\documentstyle[preprint,aps,version2]{revtex}
\begin{title}
Shock waves in one-dimensional Heisenberg ferromagnets.
\end{title}
\begin{document}
\author{V. V. Konotop$\dag$ \thanks{
Also at Center of Mathematical Sciences, University of Madeira,
Pra\c{c}a do Munic\'{\i}pio, P-9000 Funchal, Portugal.}, M.
Salerno$\ddag$ \thanks{ Also at Istituto Nazionale di Fisica
della Materia (INFM), Unita' di Salerno, I-84100
Salerno,Italy.}and S.Takeno$\S$}
\begin{instit}
$\dag$ Department of Physics, University of Madeira, Pra\c{c}a
do Munic\'{\i}pio, 9000 Funchal, Portugal and Center of Science
and Technology of Madeira (CITMA), Rua da Alf\^andega, 75-5$^o$,
Funchal, P-9000 Portugal
\end{instit}
\begin{instit}
$\ddag$ Department of Physical Sciences "E.R. Caianiello",
University of Salerno,\\ I-84100, Salerno, Italy
\end{instit}
\begin{instit}
$\S$ Faculty of Information Science, Osaka Institute of
Technology,\\ 1-79-1 Hirakata, Osaka 573-01, Japan
\end{instit}

\begin{abstract}
We use the $SU(2)$ coherent state path integral formulation with
the stationary phase approximation to investigate, both
analytically and numerically, the existence of shock waves in
the one-dimensional Heisenberg ferromagnets with anisotropic
exchange interaction. As a result we show the existence of shock
waves of two types, "bright" and "dark", which can be interpreted 
as moving magnetic domains.
\end{abstract}

\newpage
The Heisenberg model is certainly one of the most important
models of condensed matter physics on which a large amount of
work has been done \cite{r5}. In spite of this, the study of its
properties, both at the quantum and at the classical level, is
still a non exhausted subject of ever-continuing interest. In
the one dimensional case the model is exactly solvable and it is
known that the excitation spectrum consists of quantum solitons
which in the semiclassical limit can be seen as bound states of
many magnons. On the other hand, it has recently been shown that
nonlinear lattices, besides solitons, may support other kinds of
excitation which behave at the initial stages of their evolution
like shock waves in liquids or gases. Such waves have been
reported both in integrable \cite{toda,kam,kaup} and
nonintegrable \cite{holi,malomed,FPU,KS} lattices.

The aim of the present paper is to show the existence of shock
waves in the one-dimensional Heisenberg ferromagnet. To this end
we use the $SU(2)$ coherent state
\cite{Bis} path integral formulation with
the stationary phase approximation \cite{KT}, to derive
classical equation of motion out of the original quantum spin
model.  Within the framework of this approximation the system is
shown to be described by a discrete nonlinear Schr\"{o}dinger
(DNLS) like equation which has a hamiltonian structure and a non
standard Poisson bracket. Moreover, the classical system
preserves the conservation of the analogue of the quantum total
spin (for isotropic case) operator as well as of its
z-projection.  We find that for suitable condition the
excitations of this system naturally display shock waves with
sharp rectangular profiles moving on uniform backgrounds. Such
waves can exist both above background (bright shock) and below
background (dark shock) and on the contrary of other
excitations, which may decay into soliton trains or background
radiation, they are very stable. We remark that similar
solutions were found also in a deformable DNLS system \cite
{KS,S} and in a chain of two level atoms describing the
propagation of Frenkel excitons
\cite{KST}. We give a numerical and an analytical
description of these phenomena both in terms of dispersion
relations and in terms of a small amplitude multiscale
expansion.

The quantum Heisenberg Hamiltonian is written as
\begin{eqnarray}
H=-\sum_{\langle m,n\rangle }J(n,m)[(\hat S_n^x\hat S_m^x+\hat
S_n^y\hat S_m^y)+\lambda \hat S_n^z\hat S_m^z] \label{e1}
\end{eqnarray}
where $\hat S_n=(\hat S_n^x,\hat S_n^y,\hat S_n^z)$ are spin
operators of spin magnitude $S$, $J(n,m)$ is the exchange
interaction constant and $\lambda$ is the 
anisotropy of the exchange $XY$-like $(\lambda <1)$ and
Ising-like $(\lambda >1)$ interactions. In what follows we
consider the case of nearest neighbor interaction $J(n,m)=J\cdot
(\delta_{n,m+1}+\delta_{n,m-1})$ with $J>0$ and we denote, as
usual, with $\hat S_n^{\pm }=\hat S_n^x\pm i\hat S_n^y$ 
the raising and lowering operators. To derive classical
equation of motion it is suitable to use $SU(2)$ coherent states
\begin{equation}
|\mu_n\rangle =\frac{\exp (\mu_n\hat S_n^{+})}{\sqrt{1+|\mu
_n|^2}} |\downarrow \rangle_n\; \label{e2}
\end{equation}
in terms of which we write the state of the system (\ref{e1}) in
the form

\begin{equation}
|\Lambda \rangle =\prod_n^{}|\mu_n\rangle .  \label{ee3}
\end{equation}

Here $\mu_n$ is a complex variable and $|\downarrow\rangle_n$
denotes the ground state of a single site. The time evolution
operator of the Heisenberg hamiltonian between the initial state
$|\Lambda \rangle_i$ at time $t_i$ and a final state 
$|\Lambda\rangle_f$ at time $t_f$ can be
written in terms of path integral as
\begin{equation}
_i\langle \Lambda |\exp (-i\frac{H\,t}\hbar )|\Lambda \rangle
_f=\int
\partial (\Lambda )\exp (\frac i\hbar S)  \label{ee4}
\end{equation}

where $S=\int_{t_i}^{t_f}Ldt$ with $L$ given by

\begin{equation}
L=\frac{i\hbar }2\sum_n\frac 1{1+|\mu_n|^2}\left( \bar \mu_n
\frac{d\mu_n}{d\,t}-\mu_n\frac{d\bar \mu_n}{d\,t}\right)-
\langle \Lambda |H|\Lambda\rangle.  \label{ee5}
\end{equation}
Using the stationary phase approximation $\delta S=0$ in
(\ref{ee4}) one readily obtains the following equation of motion

\begin{eqnarray}
i\hbar \frac{d\mu_n}{dt} &=&-J\left( \frac{\mu_{n+1}-\mu_n^2
\bar \mu_{n+1}}{1+|\mu_{n+1}|^2}+\frac{\mu_{n-1}-\mu_n^2\bar 
\mu_{n-1}}{1+|\mu_{n-1}|^2}\right) \nonumber\\
&&+\lambda J\left(\frac{\mu_n(1-|\mu_{n+1}|^2)}{1+ |\mu_{n+1}|^2}
+\frac{\mu_n(1-|\mu_{n-1}|^2)}{1+|\mu_{n-1}|^2}\right)
\label{e4}.
\end{eqnarray}

It is of interest to note that (\ref{e4}) and its complex
conjugated follow from Hamilton's equations
\begin{equation}
\frac{d\mu_n}{dt}=\left\{ \mu_n,H_c\right\} ,\;\frac{d\bar \mu_n}{dt}
=\left\{ \bar \mu_n,H_c\right\}
\end{equation}
with the noncanonical Poisson bracket
\begin{equation}
\left\{ f,g\right\} =-\frac i\hbar \sum_n(1+|\mu_n|^2)^2\left[ \frac{
\partial f}{\partial \mu_n}\frac{\partial g}{\partial \bar \mu_n}-\frac{
\partial f}{\partial \bar \mu_n}\frac{\partial g}{\partial \mu_n}\right]
\end{equation}
and with the classical hamiltonian $H_c$ given by $\langle
\Lambda |H|\Lambda
\rangle $ i.e.
\begin{equation}
H_c=-J\sum_n\left( \frac{\bar \mu_{n+1}\mu_n+\bar \mu_n
\mu_{n+1}}{\left( 1+|\mu_n|^2\right) \left( 1+|\mu
_{n+1}|^2\right) }+\frac \lambda 2\frac{
\left( 1-|\mu_n|^2\right) \left( 1-|\mu_{n+1}|^2\right) }{\left( 1+|\mu
_n|^2\right) \left( 1+|\mu_{n+1}|^2\right) }\right) .
\end{equation}
One readily checks that the conservation of the z-component of
the quantum total spin is reflected in the classical system
(\ref{e4}) in the conservation of the quantity
\begin{equation}\label{sz}
s_z \equiv \langle \Lambda |\sum_n\hat S_n^z|\Lambda \rangle
=\frac 12
\sum_n\frac{\left( 1 -|\mu_n|^2\right) }
{\left( 1+|\mu_n|^2\right) }.
\end{equation}
In the isotropic case ($\lambda=1$) besides (\ref{sz}) there is
another conserved quantity which is just the analogue of the
total spin of the quantum system
\begin{equation}
s^2 \equiv \langle \Lambda |\hat S^2|\Lambda \rangle =
\sum_{n,m} \frac {\bar \mu_n \mu_m }{\left(1+|\mu_n|^2\right)
\left(1+|\mu_m|^2\right)}
\label{s2}
\end{equation}
[in an infinite chain with $|\mu_n|\rightarrow\rho$ as
$n\rightarrow\pm\infty$, expressions (\ref{sz}) and (\ref{s2})
should be properly normalized to be finite quantities]. The
classical system (\ref{e4}) keeps, therefore, some important
symmetry of the original quantum model [note also that the
dynamical variables $(\mu_n,\bar\mu_n)$ can be related by an
inverse stereographic projection to the motion of vectors on a
sphere (classical spins)].  In order to investigate shock
solutions of (\ref{e4}) it is suitable to consider solution
propagating against a nonzero background of the form
$\mu_n^{(0)}=\rho \exp(-i\omega t+ikn) $ 
where $\rho <1$ and
\begin{equation}
\omega =J(\lambda-2 \cos (k))\frac{1-\rho^2}{1+\rho^2}.
\label{e6}
\end{equation}
(here and in the following we fix $\hbar =1$). The stability of
the background can be studied with help of the substitution
 $\mu_n=(1+\psi_n)\rho \exp(-i\omega t+ikn)$, 
where $|\psi_n|\ll |\mu_n|$, in
(\ref{e4}). By linearization we obtain the dispersion relation
$\Omega (K)$ associated with the linear equation for 
$\psi_n$ $[\propto \exp (-i\Omega t+iKn)]$
\begin{eqnarray}
\Omega  =2\frac{1-\rho ^2}{1+\rho ^2} J\sin (k)\sin (K)
\pm \frac{2\sqrt{2}}{1+\rho ^2}J\sin \left(  \frac K2\right)\times
\nonumber \\ \left\{
\cos
^2(k)(1+\rho ^2)^2-\cos ^2(k)\cos (K)(1-\rho ^2)^2- 4\lambda
\rho ^2\cos
(k)\cos (K)\right\} ^{1/2}.\label{e8}
\end{eqnarray}
Thus the background is stable (i.e. $\Omega $ is real at all
$K$), if 
$\cos k>\lambda$ and $0> \cos
k>-\frac{2\rho ^2 }{1+\rho ^4}\lambda$.  
Naturally, in what follows the analysis will be
restricted to this region of the parameters. In order to get the
equation governing the initial stages of the evolution of a
shock wave we use the small amplitude expansion
$\mu_n=(\rho +a_n)\exp [i(-\omega t+kn-\phi
_n)]$,  
where the two real quantities $a_n$ and $\phi_n$ 
are considered depending on slow variables
$X=\gamma n$, $T=\gamma t$, and $\tau =\gamma ^3t$, (with
$\gamma \ll 1$) and are represented in the form
$a_n=\gamma ^2a_n^{(0)}+\gamma
^4a_n^{(1)}+...,$ $\phi_n=\gamma
\phi_n^{(0)}+\gamma ^3\phi_n^{(1)}+...$ .
Collecting all the terms of the same order in $\gamma$ 
we arrive at a series of equations. In the
zero order we recover the dispersion relation (\ref{e6}).  In
the second and third orders we get the following equations
\begin{eqnarray}
\frac{\partial \phi ^{(0)}}{\partial T}=\frac{8\rho J}{(1+\rho ^2)^2}(
\cos k-\lambda)a^{(0)}-2 J\sin (k)\frac{1-\rho ^2}{1+\rho
^2}\frac{\partial
\phi ^{(0)}}{\partial X},  \label{k8}
\end{eqnarray}
\begin{eqnarray}
\frac{\partial a^{(0)}}{\partial T}=\rho J\cos (k)\frac{\partial ^2\phi
^{(0)}}{\partial X^2}-2\sin (k) J\frac{1-\rho ^2}{1+\rho
^2}\frac{
\partial a^{(0)}}{\partial X}.  \label{k9}
\end{eqnarray}

It is suitable to introduce new variables ($\xi_{\pm },T$)
instead of ($X,T$), where $\xi_{\pm }=X-c_{\pm }T$ and the
velocities $c_{\pm }$ are given by
\begin{equation}
c_{\pm }=\frac{2J}{(1+\rho ^2)}\left[ (1-\rho ^2)\sin k\pm \rho
\sqrt{
2\cos k(\cos k-\lambda)}\right] .  \label{k10}
\end{equation}
Comparing this result with (\ref{e8}) one sees that $c_{\pm
}=d\Omega_{\pm }/dK$ at $K=0$, i.e. $c_{\pm }$ are group
velocities of two branches of the spectrum in the center of the
BZ. Then it follows from (\ref{k8}), (\ref{k9}) that solutions
$a^{(0)}=a^{(0)}(\xi_{\pm })=a_{\pm }$ and $\phi ^{(0)}=\phi
^{(0)}(\xi_{\pm })$ are related by
\begin{equation}
a_{\pm }=\mp (1+\rho ^2)\frac{\sqrt{\cos k}}{2\sqrt{2(\cos
k-\lambda)}}
\frac{\partial \phi ^{(0)}}{\partial \xi_{\pm }}.  \label{k11}
\end{equation}
For brevity we drop out the explicit form of the equations
appearing in the forth and fifth orders of $\gamma $ and present
simply the condition of their compatibility. This condition has
the form of a Korteweg - de Vries (KdV) equation
\begin{equation}
\frac{\partial a_{\pm }}{\partial  \tau }+\alpha (k)a_{\pm
}\frac{\partial a_{\pm }}{\partial \xi_{\pm }}+ \beta
(k)\frac{\partial ^3a_{\pm }}{\partial
\xi_{\pm }^3}=0  \label{k12}
\end{equation}
with
\begin{eqnarray}
\alpha (k) &=&\frac{4J}{(1+\rho ^2)^2}\left[ -5\rho \sin k+2\lambda\tan
k\right.  \nonumber \\ &&\ \left. \pm (1-3\rho ^2)\sqrt{2 \cos
k(\cos k-\lambda)}\pm
\frac{\lambda\sqrt{2 \cos ^3k}}{2\sqrt
{\cos k-\lambda}}(3-\rho ^2)\right]
\label{k13}
\end{eqnarray}
and
\begin{eqnarray}
\beta (k) &=&\frac J{4\rho (1+\rho ^2)\sqrt{\cos k-\lambda}}\left\{
\frac83
\rho \sin k\sqrt{\cos k-\lambda}(1-\rho ^2)\right.   \nonumber
\\
&&\ \left. \pm \sqrt{2 \cos k}\left[ \frac 23\rho ^2( \cos
k-\lambda)-4\rho ^2\lambda- \cos k(1-\rho ^2)^2\right] \right\}.
\label{k14}
\end{eqnarray}
From (\ref{k12}),(\ref{k14}) it follows that if
\begin{equation}
4\sqrt{2}\rho (1-\rho ^2)\sin k\sqrt{ \frac{\cos k-\lambda}{\cos
k}}=\pm 3[ (1-\rho ^2)^2\cos k+4\lambda \rho ^2]\mp 2\rho
^2(\cos k-\lambda)
\label{k15}
\end{equation}
is satisfied, the coefficient $\beta (k)$ becomes zero and the
KdV equation reduces to the well-known equation
\begin{equation}
\frac{\partial a_{\pm }}{\partial \tau }+\alpha (k)a_{\pm }\frac{\partial
a_{\pm }}{\partial \xi_{\pm }}=0 \label{z1}
\end{equation}
which support shock solutions. This implies that for parameter
values satisfying (\ref{k15}) shock wave should develop in the
spin chain.

There are two facts concerning (\ref{z1}) to be mentioned here.
Note that from expression (\ref{z1}) it follows that at $k=0$
there exists only one background $0<\rho<1$ at which $\beta (k)$
equals to zero while this is not true for $k\neq 0$.  Moreover,
equation (\ref{z1}) is not satisfied for all $\lambda$ values
but there exists a maximal value $\lambda_{max}=1/7$ above which
(\ref{z1}) does not have physical meaning.

To check the above predictions, we have numerically integrated
(\ref{e4}) on a long chain (we neglect boundary conditions),
taking as initial condition a bell shaped bright or dark pulse
of the type
\begin{equation}
\mu_n=\rho e^{ik n} \left(1\pm \frac A{\cosh [(n-n_0)]^2}\right)
\label{ic}
\end{equation}
(note that with this initial condition rectangular shock
profiles should develop).  In Fig. 1 we have reported the
profile which develop out from an initial bright pulse of
amplitude $|A|=3.6$, after an evolution time of $T=420$. The
background is moving in-phase ($k=0$) with $\rho=0.8$ and for
parameter values given by $J=0.8$, and $\lambda$ derived from
(\ref{k15}).  From this figure we see the appearance of a
leading rectangular shock profile followed by solitons and
background radiation. The shock wave connects the uniform
background field with a local plateau with two sharp transitions
at the edges.  If we define the local magnetization as
$M_n=\left( 1-|\mu_n|^2\right)/
\left( 1+|\mu_n|^2\right)$ 
we have that the local magnetization in the
rectangular shock waves of Fig.1 does not change in time and is
different from the surroundings.  This suggest the
interpretation of such solutions as propagating magnetic
domains.

A similar result can be obtained starting from an initial dark
profile as shown in Fig.2.  In this case the shock plateau
develops below the background and therefore it can be referred
to as a dark shock. Notice that in the above context dark and
bright pulses correspond respectively to domains with higher and
lower magnetization compared with the magnetization of the
background.

Following the time evolution of the shock profiles in Fig.s 1,2
we find that the rectangular waves separate from the other
components (solitons and radiation) and stay stable over long
time.  We have numerically checked that (\ref{k15}) is a
necessary condition for creation of shock waves.  In Fig.3 we
have reported the evolution profile for the same parameter
values of Fig.1 except for $\lambda=0.185$ not satisfying
relation (\ref{k15}).  We see that the rectangular shock is
destroyed and oscillations develop on the 
wavefront.  The same is observed for other choices of parameters
for which Eq. (\ref{k15}) is not satisfied. We also checked
that shocks remain stable upon collision with other excitations
(the same was found in ref. \cite{KS}) this suggesting in them
the presence of a strong soliton component. This leads to the
interpretation of shocks as bound states of many solitons (in
the quantum system they should correspond to long "string"
excitations i.e. to bound states with a large number of
quasiparticles). To check this interpretation, however, further
investigations are required.

In conclusion, we have shown both analytically and numerically
the existence of shock waves in 1D Heisenberg ferromagnets.
Physically, this is a novel example of manifestation of
classical fluid dynamics in quantum magnetic systems.
Previously, classical behaviors in magnetic systems were
reported on multi-magnon instabilities and chaos in pure and
doped YIG and some antiferromagnets \cite{Nak}.  In order to
observe such shock waves as those obtained here in magnetic
systems, generation of macroscopic number of magnons is
generally required.  One of promising candidates to realize this
may be highly pummped YIG.

The work of VVK has been supported by FEDER and by the Program
PRAXIS XXI, grant No.  PRAXIS/2/2.1/FIS/176/94.  MS wishes to
acknowledge financial support from INFM (Istituto Nazionale di
Fisica della Materia) and from INTAS grant 93-1324.

\noindent

\figure {\ Evolution of a bright shock against nonzero background
with $k=0$ $\rho=0.8$, and for parameter values $J=0.8$ and
$\lambda$ determined from (\ref{k15}).}\label{one}

\figure {\ Same as in Fig.1 but for a dark initial condition and for
parameter values $\rho=0.8$ $J=1$ and $\lambda$ determined from
(\ref{k15}).}
\label{two}

\figure {\ Evolution profile for the same 
parameter values as in Fig.1 but with $\lambda=0.185$ not
satisfying Eq.(\ref{k15}).}
\label{three}


\begin{references}
\bibitem{r5} For a review see for example D. C. Mattis, {\em Theory of
Magnetism, Statistics and Dynamics}, Springer Series in Solids
State Physics {\bf 17} (Springer-Verlag, Berlin, 1981) and
references therein.

\bibitem{toda}  B. L. Holian and G. K. Straub, Phys. Rev. B {\bf 18} 1593
(1978).

\bibitem{kam}  S. Kamvissis, Physica D {\bf 65}, 242 (1993)

\bibitem{kaup}  D. J. Kaup, Physica {\bf 25} D 361 (1987)

\bibitem{holi}  B. L. Holian, H. Flaska, and D. W. McLaughlin, Phys. Rev. A
{\bf 24} 2595 (1981)

\bibitem{malomed}  J. Hietarinta, T. Kuusela, and B. A. Malomed, J. Phys. A:
Math. Gen. {\bf 28}, 3015 (1995)

\bibitem{FPU}  P. Poggi, S. Ruffo, and H. Kantz, Phys. Rev. E {\bf 52}, 307
(1995)

\bibitem{KS}  V. V. Konotop and M. Salerno Phys. Rev. E (1997)

\bibitem{Bis} R. Balakrishnan, J. A. Holyst, and A. R. Bishop, J.
Condens. Matter. {\bf 2}, 1869 (1990), J. A. Holyst and L. A.
Turski, Phys. Rev. B {\bf 34}, 1937 (1986), Phys. Rev. A {\bf
45}, 6180 (1992).

\bibitem{KT}  J. W. Negele and N. Orland, {\em Quantum Many-Particle
Systems} (Addison-Wesley Pub. Com., New York, 1988) Chap. 7

\bibitem{S}  M. Salerno, Phys. Rev. A {\bf 46} 6856 (1992)

\bibitem{KST}  V. V. Konotop, M. Salerno and S. Takeno, Phys. Rev. B (1997)

\bibitem{Nak} K. Nakamura, {\em Quantum chaos - a new paradigm of
nonlinear dynamics} (Cambridge University Press, Cambridge,
U.K., 1993) and references therein.

\end{references}
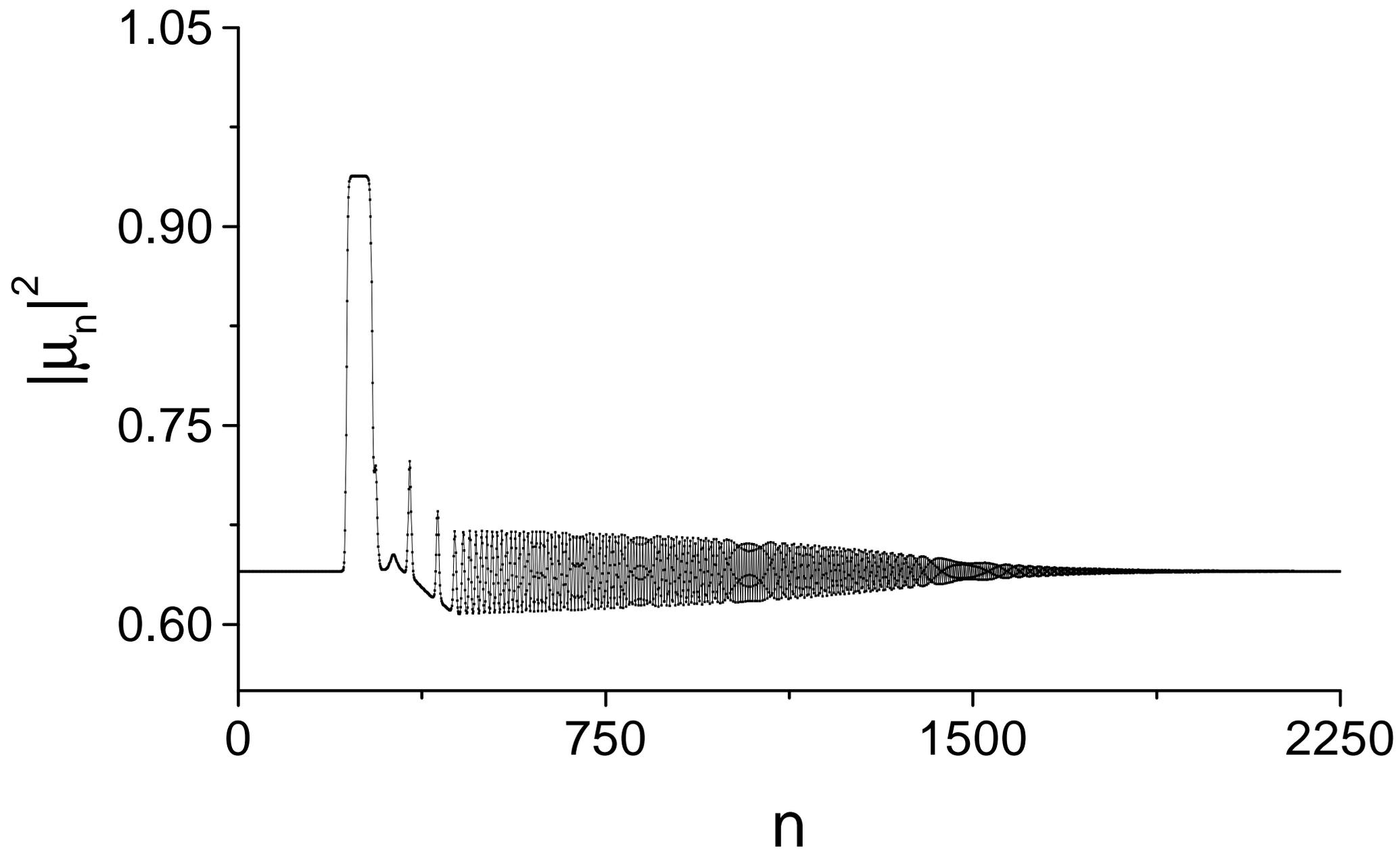
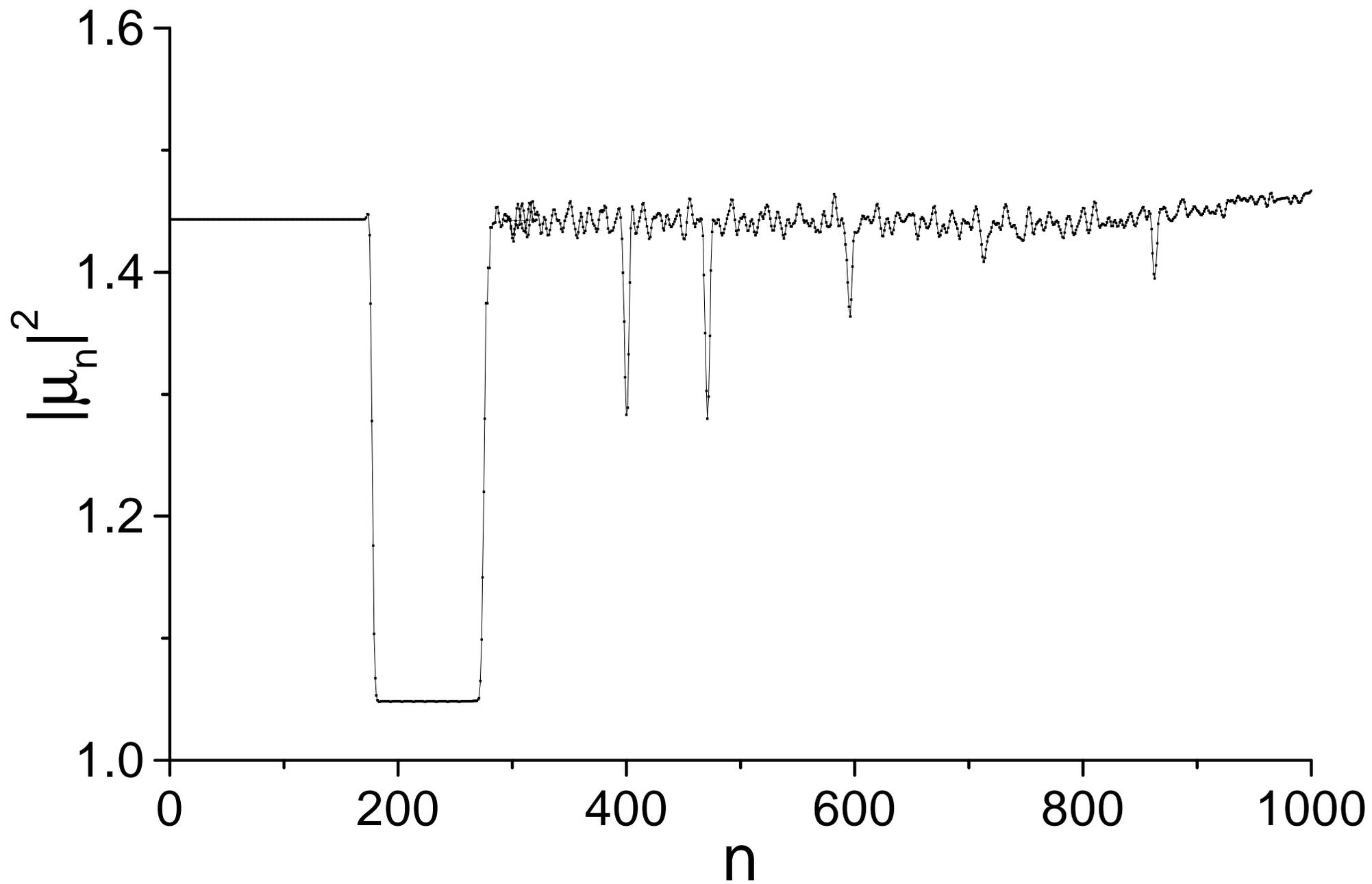
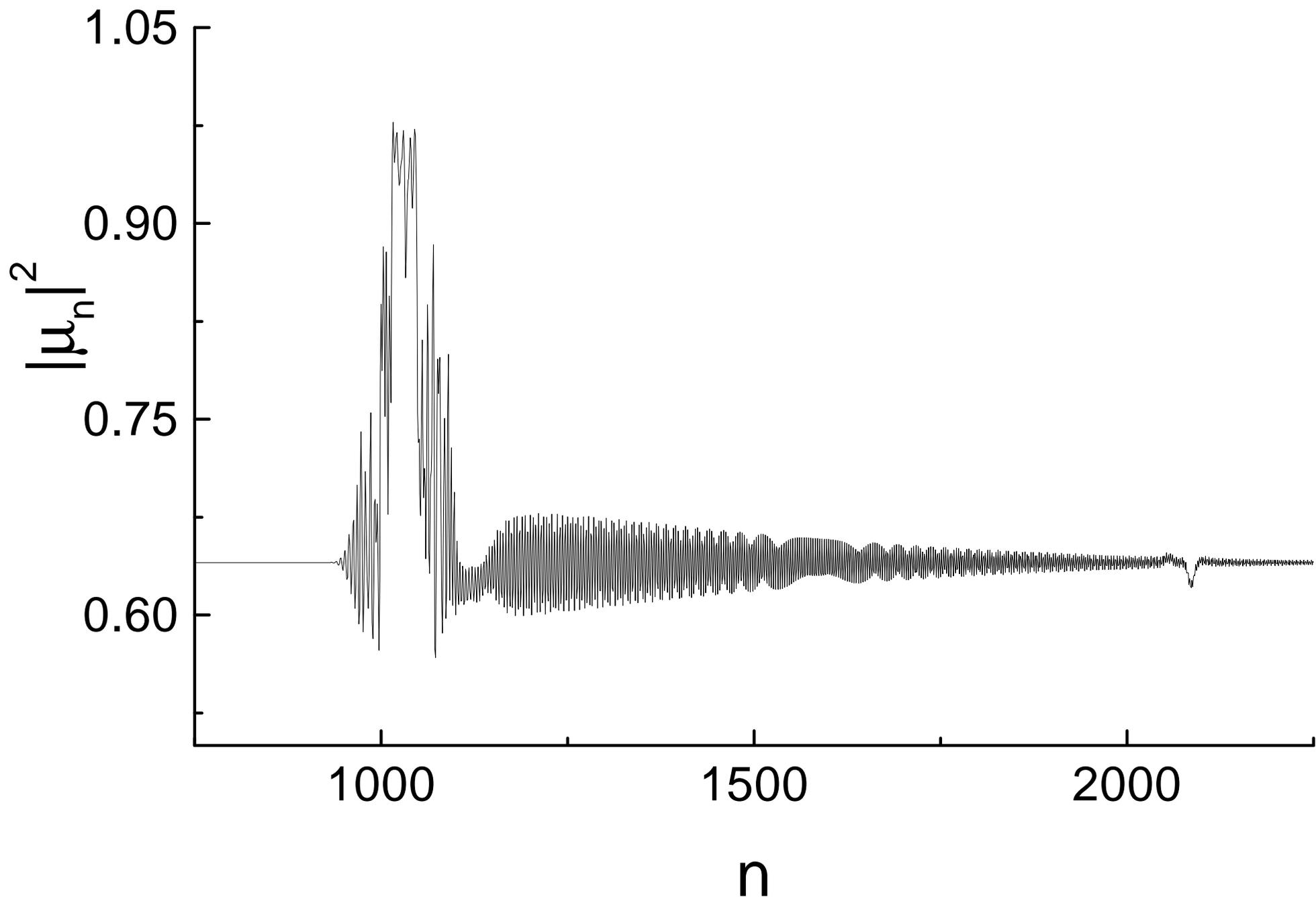
\end{document}